\documentclass[10pt,journal,compsoc]{IEEEtran}

\usepackage{graphicx}
\usepackage{multirow}
\usepackage{booktabs}
\usepackage{bigstrut}
\usepackage{hyperref}

\usepackage{graphicx}
\usepackage{url}
\usepackage{pifont}

\usepackage{subfig}
\usepackage{url}
\usepackage{threeparttable}
\usepackage{mathrsfs}

\usepackage{amsmath,amsfonts}
\usepackage{bbding}
\usepackage{titlesec}

\makeatletter  
\newif\if@restonecol  
\makeatother

\usepackage[linesnumbered,ruled,vlined]{algorithm2e}

\usepackage{algpseudocode}  
\usepackage{amsmath}  

\begin{document}
\date{Oct. 2023}
\title{DxHash: A Scalable Consistent Hashing Based on the Pseudo-Random Sequence}

\author{
{\rm Chaos Dong, Fang Wang, Dan Feng}\\
Huazhong University of Science and Technology \\ 
Emial: chaosdong@hust.edu.cn}
\maketitle	

\thispagestyle{empty}

\subsection*{Abstract}

	Consistent hashing (CH) has been pivotal as a data router and load balancer in diverse fields, including distributed databases, cloud infrastructure, and peer-to-peer networks. However, existing CH algorithms often fall short in simultaneously meeting various critical requirements, such as load balance, minimal disruption, statelessness, high lookup rate, small memory footprint, and low update overhead. To address these limitations, we introduce DxHash, a scalable consistent hashing algorithm based on pseudo-random sequences. To adjust workloads on heterogeneous nodes and enhance flexibility, we propose weighted DxHash. Through comprehensive evaluations, DxHash demonstrates substantial improvements across all six requirements compared to state-of-the-art alternatives. Notably, even when confronted with a 50\% failure ratio in a cluster of one million nodes, DxHash maintains remarkable processing capabilities, handling up to 13.3 million queries per second.

\section{Introduction}

	When the concept of Consistent Hashing (CH) was first proposed \cite{CH}, it rapidly gained popularity and assumed an essential role as a data router and load balancer in various fields, including distributed databases \cite{THDPMS,DistCache,LightStore}, cloud infrastructure \cite{SACH,ATCS}, and peer-to-peer networks \cite{Chord,AlDHT}. In these scenarios, CH evenly distributes loads (with keys) to buckets (or nodes) while maintaining consistency. Unlike conventional hashing functions, CH adheres to four notions of consistency: \emph{Balance}, \emph{Monotonicity}, \emph{Spread}, and \emph{Load} \cite{CH}. This paper assumes that node states are centrally managed by a controller, and each node holds the complete latest view of node states \cite{Beamer}. Consequently, the notions of \emph{Spread} and \emph{Load}, which come into effect in distributed hashing tables, aren't discussed in this paper. Similar to some previous work \cite{AnchorHash}, as long as a hashing algorithm satisfies the properties of \emph{Balance} and \emph{Monotonicity}, it is considered consistent. Consistency is crucial to balancing load and protecting against large-scale data migration caused by topological changes in the cluster. \par
	As CH finds wide application, in addition to the requirement of consistency, there are more pressing demands that need to be met. Recently, six properties have been identified to evaluate whether a CH algorithm is ideal: \par

	\begin{itemize}
	\item \textbf{Balance}: Ensures that each active node in the cluster has an equal probability of being mapped to by a key, resulting in a uniform load distribution among nodes.
	\item \textbf{Monotonicity} (or \textbf{Minimal Disruption}\cite{maglev}): Guarantees that when a node is inserted or removed, the affected objects are remapped either from this node to other nodes or from other nodes to this node. This property avoids unnecessary remapping among unchanged nodes.
	\item \textbf{Statelessness}: A desirable property where the results of a hashing algorithm are unaffected by historical operations (e.g., queries, node additions, and node deletions). Statelessness reduces the overhead of serializing the operating order in distributed environments.
	\item \textbf{Lookup throughput}: Measures the rate at which objects are mapped to nodes, serving as a core performance indicator for CH algorithms.
	\item \textbf{Update overhead}: Refers to the overhead incurred when node states change, such as node failures/recoveries or cluster scaling/shrinking. An ideal CH algorithm should quickly update the mapping results in response to state changes to maintain efficient routing.
	\item \textbf{Memory footprint}: Considers the amount of memory consumed by the CH algorithm. A smaller memory footprint is desirable as it reduces costs and avoids performance degradation caused by excessive memory usage \cite{SACH}.
	\end{itemize}

  \begin{table*}[htbp]
    \centering
    \caption{The Comparison of DxHash and Common CH Algorithms}
      \begin{threeparttable}          

      \begin{tabular}{lcccccc}
      \toprule
            & Ring\cite{CH}  & Maglev\cite{maglev} & JCH\cite{JumpHash} & SACH\cite{SACH} & AnchorHash\cite{AnchorHash} & DxHash \\
      \midrule
      \textbf{Balance} & +     & +++     & +++   & +++   & +++     & +++ \\
      \textbf{Monotonicity} & +++     & +     & +++   & ++   & +++     & +++ \\
      \midrule
      \textbf{Statelessness} & $\sqrt{}$     & \Checkmark\kern-1.2ex\raisebox{1ex}{\rotatebox[origin=c]{125}{\textbf{--}}}   \tnote{\ding{172}}  & $\times$   & $\times$    & $\times$      & \Checkmark\kern-1.2ex\raisebox{1ex}{\rotatebox[origin=c]{125}{\textbf{--}}} \\
      \midrule
      \textbf{Lookup} & $O(log(cn))$ \tnote{\ding{173}} & $O(1)$  & $O(log(n))$ & $O(1), O(log(m))$ \tnote{\ding{176}} & $O((1+ln(\frac{n}{a}))^2)$ \tnote{\ding{177}} & $O(\frac{n}{a})$ \\
      \textbf{Update} & $O(log(cn))$ & $O(mlog(m))$ \tnote{\ding{174}} & $O(1)$ \tnote{\ding{175}} & $O(1), O(m)$ & $O(1)$ \tnote{\ding{178}} & $O(\frac{n}{n-a})$ or $O(1)$ \tnote{\ding{179}}\\
      \textbf{Memory (Bytes)} & $24cn$ & $4m$ & $O(1)$  & $4m$  & $16n$  & $\frac{n}{8}$ to $5n$\\
      \bottomrule
      \end{tabular}%

      \begin{tablenotes}
        \footnotesize 
        \item[\ding{172}] Maglev and DxHash are statelessness for removals. The removal order of nodes does not affects the lookup results of these two CH algorithms. 
        \item[\ding{173}] \emph{Karger Ring} introduces virtual nodes for balance. Constant $c$ in \emph{Karger Ring} denotes the number of the virtual nodes pointing to each physical node, and $n$ is the number of physical nodes.
        \item[\ding{174}] $m$ in Maglev is the size of the lookup table. \textbf{The value of $m$ is recommended to be a prime number greater than $100n$} for the balance and minimal disruption.
        \item[\ding{175}] The updates in JCH are limited because only the last inserted node is allowed to be removed. 
        \item[\ding{176}] The updates in SACH are limited because the total number of nodes cannot exceed the initial maximum size. Similar to Maglev, $m$ in SACH denotes the size of the lookup table which is much larger than $a$. There are two update complexity matching the two update schemes in SACH.
        \item[\ding{177}] $n$ is the upper bound of the cluster size, and $a$ is the number of active nodes.
        \item[\ding{178}] The upper bound of AnchorHash is immutable, while DxHash supports \emph{Scale} operation to double the upper bound.
        \item[\ding{179}] The update complexity and memory footprint of DxHash are determined by the detailed implementation.
        \end{tablenotes}
      \end{threeparttable}
    \label{tab:CHcomparison}%
  \end{table*}%

	Many previously proposed methods have struggled to overcome the trade-off among the six requirements mentioned above. The original Consistent Hashing (CH), named Karger Ring or Ring in this paper, achieves minimal disruption but cannot simultaneously support both uniform balance and a low memory footprint \cite{CH}. MaglevHash \cite{maglev}, proposed by Google in 2014, faces a similar situation, although its lookup performance exceeds that of Karger Ring. SACH, proposed in 2021, is a CH algorithm similar to Maglev. Despite a series of optimizations, the memory footprint and update performance are still not ideal. Jump Consistent Hash \cite{JumpHash} and AnchorHash \cite{AnchorHash} show relatively ideal performance in six properties, but the scalability of these two has respective limitations. \par

	In this paper, we propose DxHash, a scalable consistent hashing algorithm based on the pseudo-random sequence. By iteratively selecting possible nodes using a pseudo-random generator, DxHash provides nearly ideal performance that satisfies the mentioned six properties. In the evaluation, when the cluster size exceeds 1 million nodes and 50\% of the nodes fail, DxHash can still process 13.3 million queries per second. Compared to state-of-the-art works, DxHash exhibits better lookup and update performance and improved scalability, with a smaller memory footprint. Furthermore, we combine distributed storage scenarios with DxHash to propose weighted DxHash. weighted DxHash adjusts the load on arbitrary nodes to make full use of hardware resources. \par

	The rest of the paper is organized as follows. Section \ref{sec:BK&MV} introduces related works and motivation, comparing classical or state-of-the-art CH algorithms. Sections \ref{sec:DxHash} and \ref{sec: DxHash_impl} introduce DxHash. Section \ref{sec:WDxHash} introduces weighted DxHash. In Section \ref{sec:Eva}, we evaluate the performance of DxHash in comparison with existing CH algorithms. Finally, Section \ref{sec:Con} concludes the paper. Here is a special thanks to chatGPT for contributions to improving the writing of this paper.

\section{Background} \label{sec:BK&MV}

\subsection{Related Work}

	Karger Ring, the original CH scheme proposed in 1997 \cite{CH}, maps both nodes and keys into a cyclic hash space. The ring's values increase from 0 to $2^{32}$ in the clockwise direction, and each node is responsible for the keys in its assigned segment. While Karger Ring achieves \emph{minimal disruption}, ensuring only the affected keys in the segment are remapped during node addition or removal, it struggles to maintain balance due to variable segment lengths. To address this, virtual nodes are introduced, raising the memory footprint \cite{LBPofCH}. Attempts to redistribute data for load balancing \cite{Rdis} introduce extra data migration and break \emph{minimal disruption}. Moreover, the $O(\log(n))$ complexity of Karger Ring for both update and lookup raises concerns about performance. \par

	Another CH algorithm, Highest Random Weight (HRW) \cite{HRW}, ensures complete balance and monotonicity. HRW assigns a unique identifier to each node and calculates random weights for mapping new items based on a combination of the item's key and the node's ID. The node with the highest weight is selected as the mapping result. However, HRW suffers from poor scalability due to significant computational overhead, resulting in a $O(n)$ lookup complexity, making it unsuitable for large-scale clusters. \par

	MaglevHash, proposed by Google in 2016 \cite{maglev}, is a high-efficiency CH that maintains large memory tables, where keys are hashed to table indexes, and table contents are node IDs, allowing for O(1) complexity queries. However, for balance, the table size is much larger than the number of nodes, introducing significant extra memory consumption. Additionally, MaglevHash struggles with \emph{minimal disruption} and \emph{low update complexity}. \par

	Jump Consistent Hash (JCH) is a notable CH algorithm that leverages Pseudo-Random Sequence (PRS) \cite{JumpHash}. JCH calculates a pseudo-random sequence based on a key and compares it with a specified probability to determine the node to which the key belongs. Although JCH meets standard CH requirements, it does not support arbitrary node additions or removals. Changes are restricted to the tail node, or else it would disrupt the \emph{minimal disruption} property. Consequently, JCH is not suitable for scenarios involving random and frequent node updates. \par

	Recent CH proposals, such as SACH \cite{SACH} and AnchorHash \cite{AnchorHash}, bring new perspectives. SACH uses double hashing similar to Maglev, with two update algorithms, fast but unbalanced, and slow but balanced. However, SACH still faces challenges in memory footprint and update complexity, and data skew increases with failure rates. AnchorHash, while near-ideal with $O(1+\log(\frac{a}{w}))$ lookup complexity, faces issues with a fixed upper bound on the cluster size and strict statefulness, preventing concurrent updates. \par

\subsection{Motivation}
	Since the exsiting CH algorithms has their own problems, we proposed DxHash, a stateless, scalable and consistent hash which meets the six requirements almost-perfectly. Table \ref{tab:CHcomparison} shows the theoretical performance comparisons between DxHash and other CH algorithms. We can find that all aspects of DxHash are as good as or better than others, except that the lookup complexity is a bit worse than AnchorHash. However, the experimental results in Section \ref{sec:EvaFT} will show that the practical performance of DxHash is usually higher than AnchorHash due to a minor constant term in DxHash's lookup complexity. Besides, DxHash outperforms AnchorHash because: (1) DxHash supports the scaling/shrinking operations to adjust the upper bound of the number of nodes, (2) DxHash has stronger statelessness because its mapping results are independent of the removal order of nodes, and (3) the memory footprint of DxHash is only 0.8\%-31\% of that of AnchorHash. \par

\section{DxHash Algorithm} \label{sec:DxHash}
\begin{figure}[htb]
    \center{
    \includegraphics[width=8.5cm]{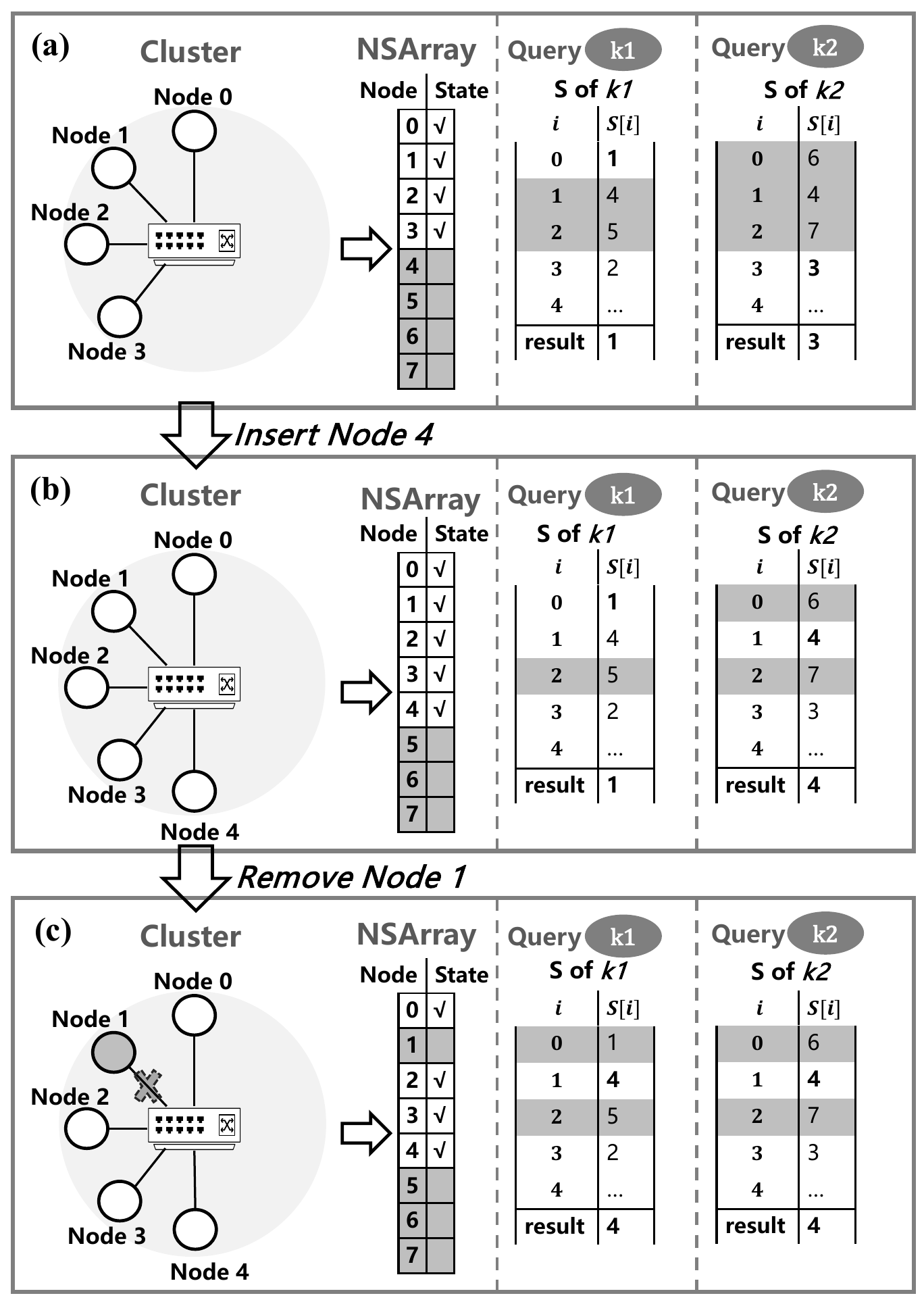}  
    \caption{An example of how DxHash works for a cluster. \textbf{(a)} Query $k1$ and $k2$. \textbf{(b)} Insert node 4. \textbf{(c)} Remove node 1.}
    \label{fig: query}}
    \end{figure}

	DxHash utilizes an array, called NSArray, to represent the state of nodes in a cluster or network. The size of the array is set to the smallest power of 2 greater than the number of active nodes in the cluster. For example, if the cluster consists of 4 nodes, the corresponding NSArray has a length of 8 (i.e., $2^3$). Each active physical node is associated with an active item in the array. Figure \ref{fig: query}a illustrates an example where the cluster contains 4 active nodes, labeled \texttt{0}, \texttt{1}, \texttt{2}, and \texttt{3}. As shown in the figure, the array items \texttt{0}, \texttt{1}, \texttt{2}, and \texttt{3} are active (depicted in white), while the remaining items in the array are reserved for future node insertions and are referred to as inactive node IDs. In the given example, the inactive items \texttt{4}, \texttt{5}, \texttt{6}, and \texttt{7} are unused and can be assigned to subsequently inserted nodes. \par

\subsection{Lookup.}
	DxHash maps objects to nodes based on a Pseudo-Random Number Generator (PRNG) \cite{PRNG2006}. A PRNG is an algorithm to generate an infinite sequence composed of pseudo-random numbers within a given range. The generated sequence is not truly random, because it is determined by an initial value called the \emph{random seed}. The PRNG features \emph{randomness} and \emph{reproducibility}. First, the generated numbers are distributed uniformly and randomly across the range. Second, the generated sequence can be reproduced if the seed is identical. \par

	By using the digitized key (e.g., the key's hash value) as the \emph{seed}, a PRNG produces a unique and reproducible sequence of random numbers for each key. By taking the items of the generated sequence modulo the size of the NSArray, DxHash obtains a sequence of node IDs that includes both active and inactive node IDs. This node-ID sequence, denoted as $S$, is used in determining the location of the key. We use $S[i]$ to denote the $i$th item in the sequence. The key's location is determined by selecting the first active item $S[i]$. Figure \ref{fig: query}a illustrates an example of querying keys \emph{k1} and \emph{k2} in a 4-node cluster. For \emph{k1}, the generated node-ID sequence $S$ is {\texttt{1}, \texttt{4}, \texttt{5}, \texttt{2}, ...}. Since the first item in the sequence, \texttt{1}, corresponds to an active node, \emph{k1} is mapped to node \texttt{1}. For \emph{k2}, $S$ is {\texttt{6}, \texttt{4}, \texttt{7}, \texttt{3}, ...}, and the first active item in the sequence is $S[3] = \texttt{3}$, resulting in \emph{k2} being mapped to node \texttt{3}.\par

	It is important to note that the node-ID sequence $S$ is calculated in real-time upon each query and is not pre-stored in memory. The number of calculations required to find an active node is referred to as the search length. When a user queries a key, they first calculate $S[0]$. If the result corresponds to an active node ID, it is the final result of the query, and the search length is 1, as shown in the example of querying \emph{k1} in Figure \ref{fig: query}a. If the result is an inactive node ID, the user continues to calculate the subsequent node IDs ($S[1]$, $S[2]$, ...) until an active node ID is found, incrementing the search length accordingly. Hence, in Figure \ref{fig: query}a, the search length for \emph{k2} is 4. \par

\subsection{Update.}
	As mentioned previously, the size of the NSArray is typically larger than the cluster size to accommodate future node insertions. When a new node joins the cluster, it is assigned an inactive node ID. In Figure \ref{fig: query}b, we illustrate an example where a new node joins the cluster and is assigned the inactive node ID 4. The corresponding item in the NSArray is then updated to active. As a result, the first active node ID in the generated node-ID sequence $S$ for \emph{k2} becomes 4, leading to the remapping of \emph{k2} to node \texttt{4}. \par

	DxHash removes nodes by marking the corresponding items in the NSArray as inactive, thus preparing them for future assignments. Figure \ref{fig: query}c demonstrates the removal of node \texttt{1}. The item \texttt{1} in the NSArray is set to inactive, causing a change in the mapping result of \emph{k1}. Initially, \emph{k1} was mapped to node \texttt{1}, but since node \texttt{1} is now inactive, \emph{k1} is remapped to node \texttt{4}. \par

\subsection{Proof of Minimal Disruption and Balance.}
In this section, we provide a proof for the properties of Minimal Disruption and Balance guaranteed by DxHash. \par

\emph{\textbf{Theorem 1 (Minimal Disruption):}} DxHash ensures Minimal Disruption. \par

\emph{\textbf{Proof:}} Let $k$ be an arbitrary key and $S$ be the generated node-ID sequence. We assume that the $n$th entry in $S$ ($S[n]$) represents the original mapping result of $k$. To prove Minimal Disruption, we consider the cases of node removal and addition: \par
($\romannumeral1$) \emph{Removal}: Suppose node $b$ is being removed. If $S[n]=b$, it implies that $k$ was originally mapped to the removed node $b$. The change in mapping does not violate Minimal Disruption. If $S[n]\neq b$, it means that $k$ was not initially mapped to node $b$. Since $\forall m < n$, the state of node $S[m]$ remains inactive and is unaffected by the removal of a node, the key $k$ continues to be mapped to node $S[n]$. \par
($\romannumeral2$) \emph{Addition}: Let $b$ be a newly added node. If $\exists l<n$ such that $b=S[l]$, we define the minimum value of $l$ as $l_{min}$. As $l_{min} < n$, for $\forall m < l_{min}$, node $S[m]$ is inactive. Consequently, $k$ will be remapped to node $S[l_{min}]=b$ because it becomes the first active node in $S$. If $\forall l < n, b \neq S[l]$, the addition of node $b$ does not impact the mapping of $k$, and therefore, the mapping of $k$ remains unchanged. \par
Thus, we can see that the changed node is either the original or the destination of the remapped keys. Consequently, DxHash achieves Minimal Disruption. \par

\emph{\textbf{Theorem 2 (Balance):}} In DxHash, there is an equal probability for a key to be mapped to each active node. \par

\emph{\textbf{Proof:}} The process of locating a key across nodes involves repeated calculations to (pseudo-)randomly generate node IDs. DxHash terminates the calculations only when a generated node ID corresponds to an active node. At the $i$th round, the probability distribution of $S[i]$ among all active nodes is uniform due to the randomness of the PRNG. Since the probability distribution at each calculation round is uniform, the overall distribution is also uniform. Thus, the Balance property is proven. \par

\subsection{Boundary cases} \label{bound case}

We have demonstrated that DxHash guarantees \emph{Minimal Disruption} and \emph{Balance}. However, there are certain boundary cases that need to be considered. In theory, with an ideal PRNG, DxHash can always access an active node as long as there exist active nodes in the cluster. However, in practice, the PRNG is not truly random and may fail to reach any active nodes. Additionally, excessively long search lengths are not acceptable. The uncontrollable search length is a unique problem for DxHash that other algorithms do not have. To bridge the gap between theory and practice, we introduce a threshold for the search length. In our implementation, the threshold is set to $8n$, where $n$ denotes the cluster size. Although setting this threshold affects balance and minimal disruption, we believe it is necessary to avoid the mentioned boundary cases. Moreover, even in scenarios where there is only one active node in a large cluster, the number of affected keys whose search length exceeds the threshold is tiny. The probability of a key not matching the only active node after $8n$ searches is:

\begin{equation}
\label{eq:EofTail}
\mathbb{P} = \left(\frac{n-1}{n}\right)^{8n}
\end{equation}

When $n$ is sufficiently large, $\mathbb{P}$ is approximately $\frac{1}{e^8}$, indicating that only 0.03\% of keys are affected by the threshold. This calculation demonstrates that terminating the search whose length is larger than $8n$ has no noticeable impact in practice. \par

One another boundary case is that, the cluster requires to scale when the cluster is full of active nodes. Some CH algorithms \cite{SACH,AnchorHash} do not support \emph{Scale} operations for two reasons. First, scaling out a cluster may require a complete remapping of keys, which can be a resource-intensive task. Second, this situation can be avoided by setting a large enough initial upper bound for the cluster. In contrast, DxHash supports \emph{Scale} operations to cater to broader applications. In DxHash, the cluster size is limited by the size of the NSArray. When the cluster reaches its maximum capacity and all items in the NSArray are active, DxHash behaves as a classic hash algorithm that maps objects to nodes with a single calculation. To scale out the cluster, DxHash doubles the size of the NSArray and sets the new items to inactive. This doubling of the range in the classic hash algorithm results in only half of the loads needing to be migrated. As a result, DxHash reduces the remapping effort by half. It's important to note that remapping half the loads can still be a significant task. Therefore, the \emph{Scale} operation is suitable for inserting active nodes in batches to amortize the overhead. For scenarios involving only a few node updates, it is recommended to initialize the NSArray with a sufficiently large size. \par

\begin{figure}[htb]
    \center{
    \includegraphics[width=8.5cm]{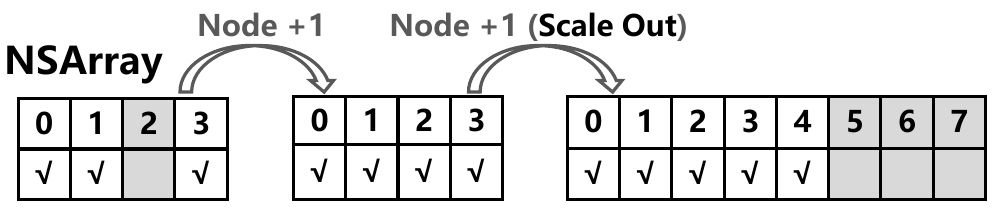}  
    \caption{An example of the scaling operation}
    \label{fig: Scaling out}}
    \end{figure}

Figure \ref{fig: Scaling out} illustrates an example of a \emph{Scale} operation. Initially, there are 3 active nodes, and the length of the NSArray is four. After inserting a new node, the NSArray becomes full, and inserting another node would require a complete remapping. To avoid this, the length of the NSArray is expanded to 8. Items 1-4 in the array are active, while the remaining items are marked as inactive, ready for subsequent node insertions. \par

\section{DxHash Implementation} \label{sec: DxHash_impl}
  
This section presents the implementation details of native DxHash, a preliminary version of DxHash. Native DxHash utilizes two data structures: NSArray and IQueue. The NSArray is implemented as a boolean array, with each item occupying only 1 bit. This bit represents the state of the corresponding node, indicating whether it is active (1) or inactive (0). On the other hand, DxHash employs a 4-byte (32-bit) integer queue called IQueue to store inactive node IDs. The functionality of IQueue will be further discussed in \S\ref{sec:addnode}. \par
Native DxHash supports four essential functions: \emph{Lookup}, \emph{AddNode}, \emph{RmNode}, and \emph{Init}. Algorithm \ref{alg:DxHash1.0} outlines the functions.

  \begin{algorithm} 
    \caption{Native DxHash}
    \label{alg:DxHash1.0}
    \SetKwProg{Fn}{Function}{:}{}
    \SetKwFunction{FLookup}{$Lookup$}%
    \SetKwFunction{FRmNode}{$RmNode$}%
    \SetKwFunction{FAddNode}{$AddNode$}%
    \SetKwFunction{FInit}{$Init$}%

    \tcc{This function receives a given key and output the corresponding node. }
    \Fn{\FLookup{$k$}}{ 
    \KwResult{A working node ID ($n_{ID}$)}
      $r \gets k$\;
      \Repeat{\texttt{NSArray}[$n_{ID}$] = 1}
      {
        $r \gets R(r)$\;
        $n_{ID} \gets r$ mod $|\texttt{NSArray}|$\;
      }
      \KwRet{$n_{ID}$} \;
    }

    \tcc{This function finds an inactive node ID and rests the ID to active. }
    \Fn{\FAddNode{}}{ 
      \KwResult{An inactive node ID ($n_{ID}$)}
      $n_{ID} \gets$ \texttt{IQueue}.pop()\;
      \texttt{NSArray}[$n_{ID}$] $\gets 1$\;
      \KwRet{$n_{ID}$}\;  
    }

    \tcc{This function receives an active node ID to remove it. }
    \Fn{\FRmNode{$n_{ID}$}}{ 
      \KwResult{Void}
      \texttt{NSArray}[$n_{ID}$] $\gets 0$ \;
      \texttt{IQueue}.push($n_{ID}$) \;
      \KwRet{} \;
    }

    \tcc{This function initializes the IQueue by a given NSArray. }
    \Fn{\FInit{$\texttt{NSArray}$}}{ 
      \KwResult{Void}

      \texttt{IQueue} $\gets$ $\emptyset$\;  
      \For{$item \in \texttt{NSArray}$}  
      {  
        \If{$item = 0$}
        {
          \texttt{IQueue}.push($item$)\;
        }
      }
      \KwRet{}\;
    }
    \end{algorithm}

\subsection{Lookup}

\emph{Lookup} is the core function of DxHash, responsible for mapping a key to its corresponding node ID. DxHash utilizes a PRNG $R(x)$ to generate pseudo-random numbers cyclically (lines 3-6), with the key serving as the random seed (line 2). By performing modulo operation with the size of the NSArray, a random node ID is obtained (line 5). The loop continues until an active node is encountered (line 6), at which point the loop terminates, and the node ID is returned as the lookup result (line 7). \par

The time complexity of the \emph{Lookup} function significantly impacts DxHash's performance. Let $n$ denote the length of the NSArray and $a$ denote the number of active nodes. We use $p$ to represent the fraction $\frac{a}{n}$, which corresponds to the active ratio in the NSArray. In each iteration of the \emph{Lookup} function (lines 3-6), the probability of hitting an active node is $p$, while the probability of hitting an inactive node is $(1-p)$. The distribution of hitting an active node at each iteration follows the Bernoulli Distribution, and the number of iterations (i.e., search length) follows the Geometric Distribution \cite{GD}. Denoting the search length as $\tau$, the expected value of $\tau$ is:\par

\begin{equation}
\label{for:EofASL}
\mathbb{E}(\tau) = \frac{1}{p} \\
\end{equation}

Substituting $p=\frac{a}{n}$ back into Formula \ref{for:EofASL}, we find: \par

\emph{Theorem 3 (Query Complexity)}: In DxHash, given the size of the NSArray, $n$, and the number of active nodes, $a$, the Average Search Length (ASL) for keys is $\frac{n}{a}$.\par

\subsection{AddNode} \label{sec:addnode}

The \emph{AddNode} function is responsible for node insertions in DxHash. When a new node joins the cluster, DxHash assigns an inactive ID to the node and adjusts the corresponding item in the NSArray. The key issue in this process is obtaining an inactive ID efficiently. Instead of performing a linear search on the NSArray, which has a time complexity of O(n), DxHash introduces a new data structure called IQueue to expedite insertions. IQueue is a 4-byte (32-bit) integer queue that stores all inactive node IDs for fast insertions. In the pseudocode of the native DxHash (Alg. \ref{alg:DxHash1.0}), lines 9-11 demonstrate how DxHash obtains an inactive node ID from the IQueue in constant time complexity (O(1)). \par


\subsection{RmNode}
The \emph{RmNode} function handles node removals. When a node is removed, DxHash updates the data structures, namely the NSArray and IQueue. In Alg. \ref{alg:DxHash1.0}, the \emph{RmNode} function receives an active node ID as input. First, the corresponding item in the NSArray is set to 0 to indicate that the node is inactive. Then, the node ID is added to the IQueue for future assignments. The time complexity of the \emph{RmNode} function is also O(1). \par

\subsection{Scale}
DxHash supports the scale and shrink operations to adjust the upper bound of the cluster size. When the NSArray is full and new nodes are ready to join, the \emph{Scale} operation is triggered. Alg. \ref{alg:adjustment} presents the pseudocode for the \emph{Scale} function. In this operation, the size of the NSArray is doubled (lines 2-3), and the new node IDs are set to be inactive (lines 4-5). \par
  
\subsection{Shrink}
The \emph{Shrink} function is used when there are too many inactive nodes in the cluster. First, DxHash counts the number of active nodes with IDs greater than $\frac{|NSArray|}{2}$ (lines 8-12, Alg. \ref{alg:adjustment}). Then, the size of the NSArray is halved (line 13), and the IQueue is rebuilt based on the updated NSArray (line 14). Finally, DxHash reassigns the same number of nodes counted in the first step (lines 15-16). \par

When scaling out or shrinking the cluster, DxHash operates on the NSArray and IQueue, involving a maximum of $n$ nodes. The time complexity of both the \emph{Scale} and \emph{Shrink} operations is O(n). However, the number of remapped keys caused by the \emph{Shrink} operation is relatively greater than that caused by the \emph{Scale} operation. This is because the \emph{Shrink} operation includes additional node deletions and insertions. The \emph{Shrink} operation is triggered only when the active ratio is very small (e.g., 1\%). \par

\begin{algorithm} 
    \caption{Scale adjustment}
    \label{alg:adjustment}
    \SetKwProg{Fn}{Function}{:}{}
    \SetKwFunction{FScale}{$Scale\_out$}%
    \SetKwFunction{FShrink}{$Shink$}%

    \Fn{\FScale{}}{ 
      \KwResult{The cluster size after adjustment}

      $n \gets |\texttt{NSArray}|$\;
      \texttt{NSArray} is resized to $2n$\;
      \For{$i\in[n, 2n)$}
      {
        RmNode($i$)\;
      }
      \KwRet{2$n$} \;
    }

    \Fn{\FShrink{}}{ 
      \KwResult{The cluster size after adjustment}
      $n \gets |\texttt{NSArray}|$\;
      $count \gets 0$\;
      \For{$i\in[\frac{n}{2}, n)$}
      {
        \If{$\texttt{NSArray}[i] = 1$}
        {      
          $count \gets count+1$\;
        }
      }
      \texttt{NSArray} is resized to $\frac{n}{2}$\;
      Init(\texttt{NSArray})\;
      \For{$i \in [0, count)$}
      {
        AddNode()\;
      }
      \KwRet{$\frac{n}{2}$} \;
    }
  \end{algorithm} 

\subsection{Optional Trade-off}

\subsubsection{New method for insertion: minimal memory footprint but slower insertions. }

DxHash introduces the IQueue data structure to accelerate insertions, but this comes at the cost of storage efficiency and statelessness. The IQueue can vary in length from 0 to $n$, and each item in the queue requires 4 bytes of storage. On average, the expected memory footprint of the IQueue is $2n$ bytes. In comparison, the size of the NSArray is much smaller, only $\frac{n}{8}$ bytes. Additionally, the use of IQueue for insertions introduces a dependency on the removal order, which compromises the statelessness property. The sequence of node IDs in the IQueue is determined by the order of removals. Consequently, the order of insertions becomes deterministic and stateful, which goes against the desired statelessness characteristic of the DxHash algorithm. To address this, we propose an alternative insertion approach for DxHash to achieve less memory and stronger statelessness without the need for an additional data structure. \par

The key to inserting a node is to obtain an inactive node ID for assignment. The new insertion approach is inspired by the \emph{Lookup} procedure in DxHash, which accesses NSArray items pseudo-randomly and repeatedly until an inactive item is found. The pseudocode for this approach is shown in Alg. \ref{alg:DxHash2.0}. In lines 2-6, we reuse the code from the \emph{Lookup} procedure with two modifications. First, we change the random seed to use a constant instead of a given key for reproducibility, which is 1228 shown as line 2. Second, we terminate the loop when an inactive node is selected instead of an active one, as we are looking for an inactive node ID (line 6). Similar to the calculation in Formula \ref{for:EofASL}, the time complexity of this insertion approach can be estimated as $O(\frac{n}{n-a})$. \par

  \begin{algorithm}
    \caption{AddNode\_2}
    \label{alg:DxHash2.0}
    \SetKwProg{Fn}{Function}{:}{}
    \SetKwFunction{FAddNode_New}{$AddNode_2$}%

    \tcc{This function finds an inactive node ID and resets the ID to active. }
    \Fn{\FAddNode{}}{ 
      \KwResult{An inactive node ID ($n_{ID}$)}
      $r \gets 1228$\;
      \Repeat{\texttt{NSArray}[$n_{ID}$] = 0}
      {
        $r \gets R(r)$\;
        $n_{ID} \gets r$ mod $|\texttt{NSArray}|$\;
      }
      \texttt{NSArray}[$n_{ID}$] $\gets 1$\;
      \KwRet{$n_{ID}$}\;
    }
  \end{algorithm}

This new insertion approach strikes a balance between space footprint, statelessness, and update efficiency. Firstly, since no additional data structure is required, the memory footprint of DxHash is solely determined by the size of the NSArray, which is $\frac{n}{8}$ bytes or 125 KB per million nodes. This memory footprint is only 0.8\% of that of \emph{AnchorHash} ($16n$ bytes). Secondly, the new design achieves stronger statelessness, as the insertion order is unaffected by the history of node removals. The shortcoming is the higher insertion complexity, which is $O(\frac{n}{n-a})$ now. Although the insertion efficiency decreases, it is still superior to most CH algorithms. The size of the NSArray $n$ is twice the number of active nodes $a$ initially. When there are only a few additions or removals, the time complexity of $O(\frac{n}{n-a})$ remains relatively constant and independent of the absolute cluster size. In the worst-case scenario where only one inactive node exists ($n-a=1$), the time complexity is $O(n)$, which is no worse than the method of linear search. It is worth noting that the cases where no inactive or active nodes exist are not considered in this section, as the boundary cases are discussed separately in \S\ref{bound case}. In summary, DxHash without IQueue is a stateless CH algorithm with a small memory footprint and acceptable update overhead. \par

\subsubsection{NSArray organized in Bytes: faster lookup but larger memory.}
The native implementation of DxHash uses 1 bit for each node to significantly reduce the memory footprint. However, since memory devices are typically byte-addressed, performing operations on bytes is much faster than on individual bits. Therefore, when high lookup performance is desired, the NSArray in DxHash can be organized as a byte array, with each node's state represented by a single byte. \par

Compared to the native implementation of DxHash, this solution sacrifices some storage efficiency in exchange for improved lookup performance. However, the memory footprint is still lower than that of \emph{AnchorHash}. As the NSArray becomes a byte array, its memory footprint is equal to $n$ bytes. Taking into account the memory usage of IQueue, which is $4a$ at most, the total memory requirement for this approach is $5n$ bytes, resulting in a 69\% reduction in memory compared to \emph{AnchorHash} ($16n$ bytes). \par

\section{Weighted DxHash} \label{sec:WDxHash}

Weighted DxHash is introduced to address the issue of load distribution in clusters or networks consisting of heterogeneous physical nodes. Conventional consistent hashing schemes use virtual nodes to adjust load distribution, where multiple virtual nodes point to the same physical node, effectively multiplying the load on that node. However, virtual nodes lead to increased memory footprint and cannot accurately distribute the load. In response, we propose Weighted DxHash. \par

\begin{figure}[htb]
\center{
\includegraphics[width=8.5cm]{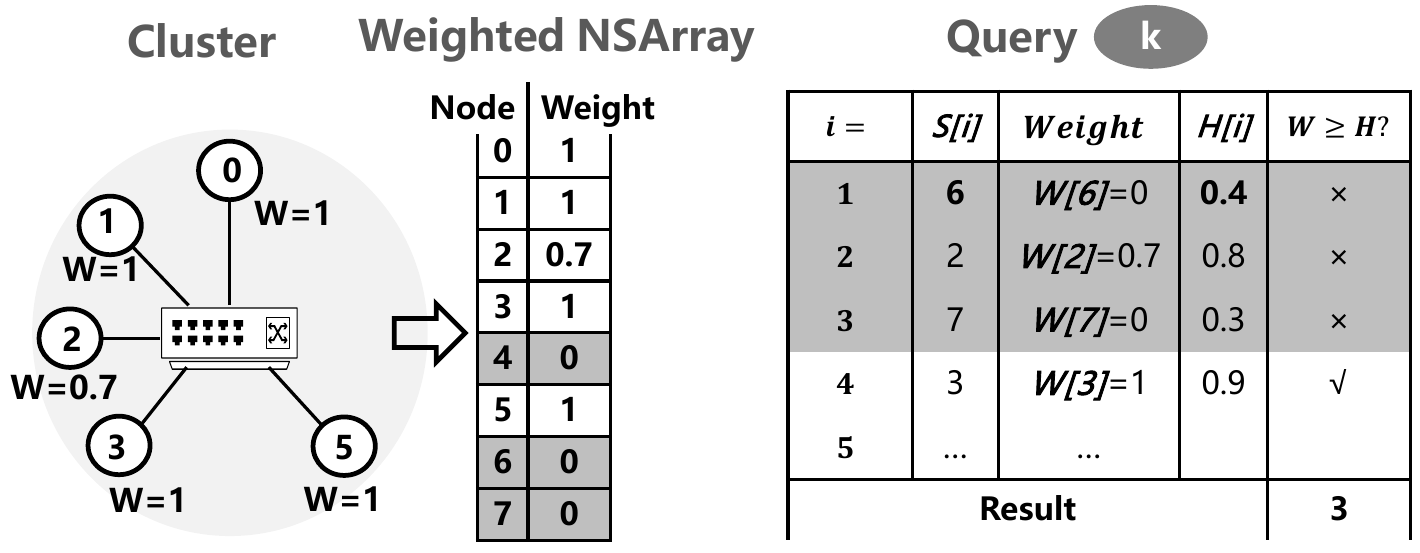}  
\caption{An example of querying a key in a 5-node weighted cluster via weighted DxHash. The mapping result of the key is node 3. }
\label{fig: weighted}}
\end{figure}

Weighted DxHash extends the native DxHash by introducing node weights and another PRNG called $H$. Each node is assigned a random floating-point weight between 0 and 1. Nodes with higher weights can handle more loads, while the weight of an inactive node is set to 0. The PRNG $H$ is used to generate a pseudo-random floating-point sequence within the range of [0, 1], with $H[i]$ representing the $i$th item in the sequence. \par

Figure \ref{fig: weighted} illustrates an example of querying a key using Weighted DxHash. The cluster shown in the figure consists of 5 nodes, with node \texttt{2} having a weight of 0.7 and all other nodes having a weight of 1. The cluster is represented as a weighted NSArray of length 8. The weights of the inactive items in the array (nodes \texttt{4}, \texttt{6}, \texttt{7}) are set to 0. The lookup process in Weighted DxHash is completed in two steps. \par

In step 1, similar to DxHash, Weighted DxHash generates a random node ID $S[i]$ at each calculation cycle. In the right part of Figure \ref{fig: weighted}, the random node IDs generated for the key $k$ are \texttt{6, 2, 7, 3, ...} in sequence. \par

In step 2, Weighted DxHash generates $H[i]$ and compares it with the weight of node $S[i]$. If the weight is no less than $H[i]$, Weighted DxHash terminates the loop and returns the current node ID as the mapping result. Otherwise, Weighted DxHash proceeds to the next cycle for further searching. From the last column in Figure \ref{fig: weighted}, we observe that in the first 3 cycles, the weights of the generated node IDs are always smaller than $H[i]$. However, at $i=4$, the weight of node $S[4]=3$ is 1, which is larger than $H[4]=0.9$. Therefore, node \texttt{3} becomes the final mapping result for the key $k$. \par

The main idea behind Weighted DxHash is to influence the probability of a key being mapped to a node based on the node's weight. In step 1 of Weighted DxHash, keys pointing to node $S[i]$ with weight $W$ have a probability of $W$ to be accepted by that node in step 2. As the weight decreases, the node becomes less likely to accept keys, resulting in a reduced load. When the weight is 0, the node rejects all keys, effectively having no load. If weights are only set to 0 and 1, Weighted DxHash is downgraded to native DxHash. \par

The time complexity of the \emph{Lookup} operation in Weighted DxHash can be analyzed as follows. The probability for a key to hit a node in round $i$ is $\frac{\sum_{x=0}^{n-1}W_x}{n}$, where $n$ is the size of the weighted NSArray and $W_x$ is the weight of node $x$. There are two calculations performed at each round, one for $S$ and another for $H$. The number of calculations required to query a key, denoted as $\tau$, follows the Geometric Distribution, and its expectation is given by: \par

\begin{equation}
\label{for:weightedASL}
\mathbb{E}(\tau)=\frac{2n}{\sum_{i=0}^{a-1}W_i} 
\end{equation} \par

The expectation of the load on node $b$, denoted as $l_b$, can be calculated using the following formula:
\begin{equation}
\label{for:weightedAcc}
\mathbb{E}(l_b)=\frac{W_b}{\sum_{i=0}^{n-1}W_i}*L 
\end{equation}
where $L$ represents the total load across all nodes. \par

In terms of space complexity, the weighted NSArray in Weighted DxHash is organized as a 4-Byte (32-bit) floating-point array. The rest of the implementation remains the same. Therefore, the expected memory footprint is $8n$ Bytes, which includes both the 4-Byte NSArray and the 4-Byte IQueue. \par

Weighted DxHash overcomes the limitations of virtual nodes and provides flexibility in load distribution. By combining virtual nodes with node weights, the load distribution can be adjusted according to the performance of individual nodes. Nodes with lower performance can be assigned smaller weights to reduce their load, while high-performance nodes can have multiple virtual nodes to fully utilize their capabilities. \par

\section{Evaluation} \label{sec:Eva}
In this section, we compare and evaluate the performance of different consistent hashing (CH) algorithms, including \emph{Karger Ring} (Ring), \emph{MaglevHash} (Maglev), \emph{AnchorHash} (AH), and DxHash (DH). We also consider different trade-offs in the implementation of DxHash, such as using a 1-bit NSArray or a 1-Byte NSArray, and using IQueue or not. These different implementations are denoted as DH-b, DH-B, and DH-IQ, respectively. Since deploying a large-scale cluster for testing is challenging, we evaluate the CH algorithms through local simulations. \par

To evaluate the performance of DxHash, we initialize an NSArray with 1 million entries, representing the states of 1 million mock nodes (active or inactive). We then generate batches of 32-bit integers as keys and feed them to DxHash. The algorithm returns the corresponding mock node IDs. We measure the lookup performance as the rate of successful key queries, the memory footprint as the amount of memory used by the process, and the update overhead as the time consumed to adjust the data structures when inserting new nodes. Similar evaluation methods are used for the other CH algorithms. \par

\subsection{Environment}

As shown in Table \ref{tab:environment}, all experiments are performed on the same commercial machine with the processor of Intel Xeon CPU E5-2620 at 2.00GHz and 32 GB memory. The system is CentOS 7.8. The kernel version is 3.10.0-1127, and the GCC version is 7.3.1. All algorithms are implemented in C++. The PRNG to generate $S$ is implemented as a hardware-supported CRC32 \cite{CRC32} hash function, which a a uniform hash function for 32-bit integers with high randomness and speed in number generation. \par

\begin{table}[htbp]
  \centering
  \caption{Environment Configuration}
    \begin{tabular*}{9cm}{ll}
    \toprule
    Processor & Intel Xeon CPU E5-2620 0 @ 2.00GHz\\
    Memory & 32GB \\
    Operating System & CentOS Linux release 7.8.2003 (Core) \\
    Kernel Version & 3.10.0-1127.13.1.el7.x86\_64 \\
    GCC Version  & 7.3.1 20180303 \\
    \bottomrule
    \end{tabular*}%
  \label{tab:environment}%
\end{table}%

In this section, we evaluate DxHash (DH) in terms of memory footprint, lookup throughput, update overhead, load balancing, minimal disruption, fault tolerance, and elasticity. \par

\begin{figure}[htb]
\center{
  \includegraphics[width=10cm]{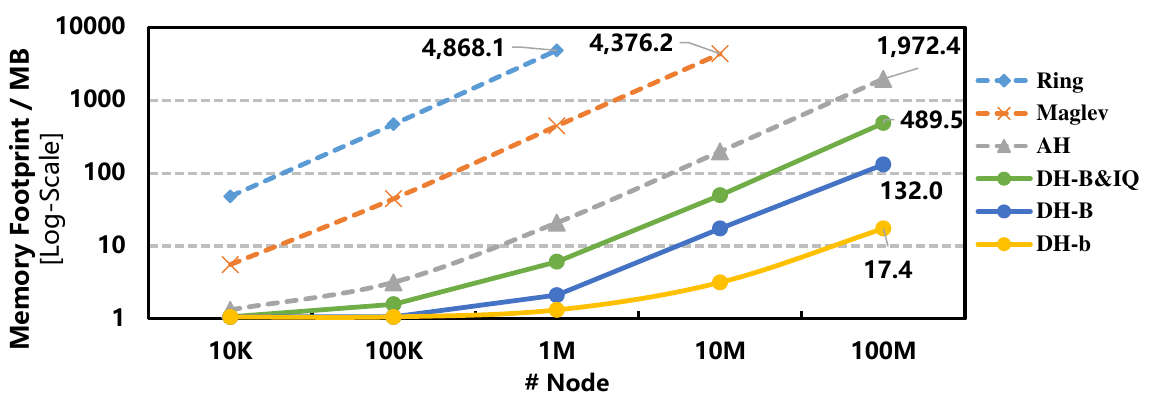}  
  \caption{Memory Footprint of Ring, Maglev, AH, DH-B\&IQ, DH-B, DH-b in the cluster whose size varies from 10K to 100M.}
  \label{fig: memory_footprint}
}
\end{figure}

\begin{figure*}[htb]
\center{
  \includegraphics[width=14cm]{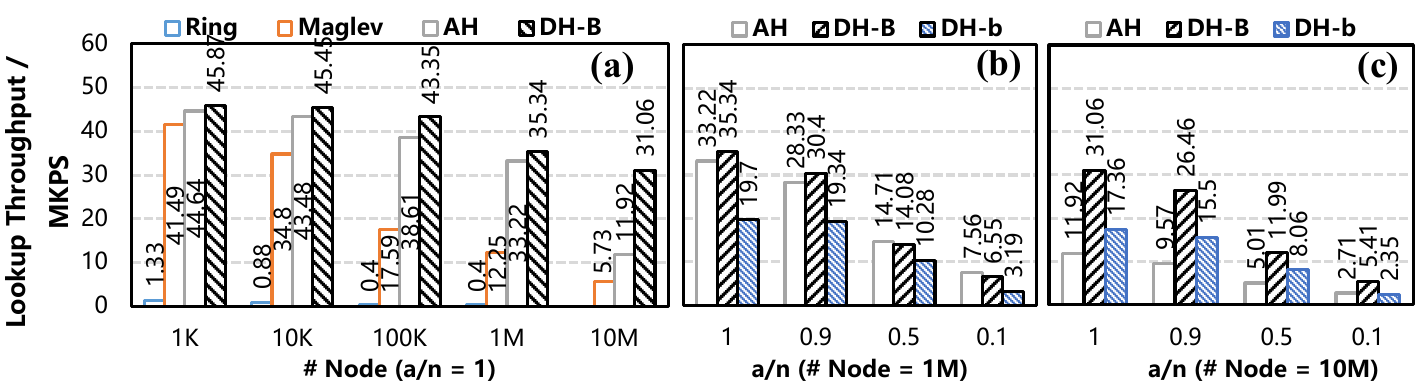}  
  \caption{The lookup rate when handling 100 million queries. (a) Lookup comparison of Maglev, Ring, AH and DH-B in clusters with 1K, 10K, 100K, 1M and 10M nodes. The active ratio of AH and DH-B ($a/n$) is 1. (b) Lookup comparison of AH, DH-B, and DH-b when the active ratio ($a/n$) varies from 1 to 0.1. The cluster size is 1 million. (c) Lookup comparison of AH, DH-B, and DH-b when the cluster size is 10 million.}
  \label{fig:lookuprate_1}
}
\end{figure*}

\subsection{Memory Footprint}
CH algorithms have varying memory footprints depending on their implementation. Ring and Maglev allocate additional memory space for load balancing or to minimize migration after node updates. Ring uses a Red-Black Tree (RBTree) implementation, with each node occupying 24 Bytes of memory. In our experiments, we create 100 virtual nodes for each physical node in Ring to achieve load balance. Therefore, Ring's memory footprint is approximately $2.4n$ KB. Maglev utilizes a large lookup table for key routing, with each entry requiring 4 Bytes of memory. We allocate 100 times the minimum required memory for Maglev to ensure less than 1\% imbalance \cite{maglev}, resulting in a memory footprint of $400n$ Bytes. AH has a memory footprint of $16n$ Bytes, while DH's memory footprint varies depending on the version. We implements three versions of DH. DH-b, which uses a bit array as NSArray, requires $\frac{n}{8}$ Bytes of memory. DH-B, which uses a byte array as NSArray, occupies $n$ Bytes. DH-B\&IQ, based on DH-B, utilizes a 32-bit IQueue to store inactive node IDs, resulting in a memory footprint of up to $5n$ Bytes. Theoretical analysis confirms that DH has a smaller memory footprint compared to other CH algorithms. \par

To validate the theoretical analysis, we collected the memory footprints of the CH algorithms for different numbers of nodes and present the results in Figure \ref{fig: memory_footprint}. The x-axis represents the number of nodes ranging from 10K to 100M. The experimental results align with our theoretical derivation. In a 100-million-node cluster, AH requires 2 GB of memory, DH-B\&IQ occupies 500 MB, DH-B requires 100 MB, and DH-b only uses 17 MB. DH-b is the most memory-saving implementations, reducing the memory footprint by 98.4\% compared to AH. DH-B\&IQ uses most memory in the three DH's implementations, still with a reduction on memory footprint of 75.2\%. \par

\begin{figure*}[htb]
\center{
  \includegraphics[width=12cm]{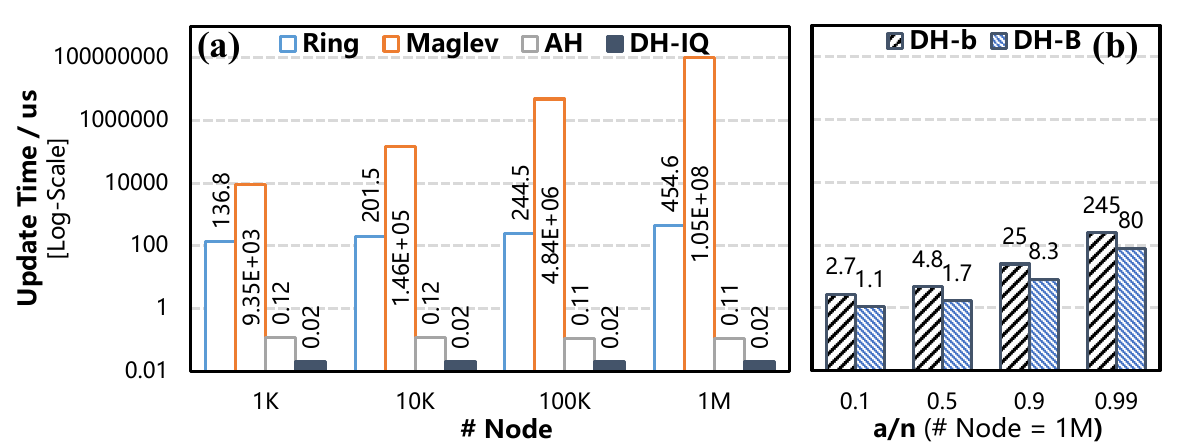}  
  \caption{Insertion overhead comparison. (a) Insertion latency of Maglev, Ring, AH and DH-IQ in clusters with 1K, 10K, 100K, and 1M nodes. (b) Insertion latency of DH-B and DH-b when the active ratio ($a/n$) varies from 0.1 to 0.99. The cluster size is 1 million.}
  \label{fig: updatetime}
}
\end{figure*}

\subsection{Lookup Throughput} \label{sec: EV_lookup}

We compare the lookup throughput of Ring, Maglev, AH, and DH. Since DH offers optional trade-offs between lookup efficiency and memory footprint, we test two versions of DH: DH-B, which uses a byte array as the NSArray for higher lookup rate, and DH-b, which uses a bit array for a smaller memory footprint. \par

Figure \ref{fig:lookuprate_1}a shows the impact of cluster scale on the lookup performance of the different CH algorithms. We vary the number of nodes from 1K to 10M and measure the lookup rate in Million Keys Per Second (MKPS) on the y-axis. In this case, all nodes are active, resulting in an active ratio ($a/n$) of 1. Figures \ref{fig:lookuprate_1}(b) and \ref{fig:lookuprate_1}(c) demonstrate how the active ratio affects the lookup performance of AH and DH (including DH-b and DH-B). The active ratio is varied between 1, 0.9, 0.5, and 0.1. Figure \ref{fig:lookuprate_1}b corresponds to a cluster with 1 million nodes, while Figure \ref{fig:lookuprate_1}c represents a cluster with 10 million nodes. From the figures, we can make several observations: \par

\ding{172} DH-B outperforms all other CH algorithms in terms of lookup throughput. AH performs slightly worse than DH-B, while Maglev has better lookup performance than Ring but is still inferior to DH-B and AH. This is because the lookup complexity of Ring is $O(\log(n))$, which is relatively worse than the $O(1)$ complexity of other CH algorithms.

\ding{173} Maglev and AH exhibit satisfactory lookup performance when the number of nodes is small. For example, when there are 1,000 nodes, Maglev achieves a lookup rate of 41.5 MKPS, and AH achieves a lookup rate of 44.64 MKPS, which is close to the lookup rate of DH-B. However, their lookup performance decreases as the number of nodes increases. When there are 10 million nodes, Maglev's lookup rate drops to only 5.73 MKPS, and AH's lookup rate drops to 11.92 MKPS. This rapid decline is due to their excessive memory footprint, which becomes a bottleneck for quick querying as the number of nodes increases. In contrast, DH-B maintains a lookup rate of 31.06 MKPS even with a 10-million-node cluster, which is 2.6 times higher than AH and 5.4 times higher than Maglev. This high lookup throughput is due to DH's nearly constant complexity and its tiny memory footprint. \par

\ding{173} Figure \ref{fig:lookuprate_1}b shows the lookup performance of AH, DH-B, and DH-b in a 1-million-node cluster with different active ratios. As the active ratio decreases, the lookup throughput of all three algorithms decreases. However, AH experiences a slower drop in performance compared to DH-B. When the active ratio is less than 0.5, DH-B performs worse than AH. This is because the lookup complexity of DH has a linear correlation with the active ratio ($O(a/n)$), while AH has a logarithmic correlation with the active ratio ($O((1+\ln(\frac{n}{a}))^2)$).

\ding{178} From Figure \ref{fig:lookuprate_1}c, we observe that although AH has better lookup complexity, it performs much worse than DH-B in a 10-million-node cluster. The lookup throughput of DH-B is $2-2.7\times$ higher than that of AH. This can be attributed to DH's smaller memory footprint, which allows it to be stored in the CPU cache for faster access. Additionally, DH has a simpler lookup method that requires fewer memory accesses compared to AH, which is discussed in Section \ref{sec:EvaFT}. The smaller and simpler data structure in DH incurs lower memory access costs compared to the larger and more complex data structure in AH.\par

Overall, DH-B demonstrates comparable lookup throughput with AH and outperforms other CH algorithms, particularly in large-scale clusters. Its high performance can be attributed to its nearly constant complexity, minimal memory footprint, and efficient lookup method. \par

\subsection{Update overhead}

Update overhead is another critical metric for evaluating CH algorithms. Figure \ref{fig: updatetime} illustrates the time required for inserting nodes into different CH algorithms. DH is implemented in three versions based on optional trade-offs. DH-IQ uses the IQueue for fast insertions, while DH-B and DH-b employ a slower method that accesses NSArray repeatedly and randomly to select an inactive node ID for assignment. Although the latter method is slower, it reduces memory footprint significantly. The efficiency of this method is related to the active ratio, where smaller active ratios result in faster insertions. The difference between DH-B and DH-b lies in the data structure of NSArray, with DH-B using a byte array and DH-b using a bit array. \par

In Figure \ref{fig: updatetime}(a), we observe the influence of cluster scale on the update cost of CH algorithms. The x-axis represents the node number, ranging from 1K to 1M, while the y-axis represents the time taken to adjust the CH algorithm after inserting a new node. The results are averaged over 100 trials. Figure \ref{fig: updatetime}(b) displays the update cost of DH-B and DH-b for different active ratios. We make the following observations from the figures: \par

\ding{172} Maglev exhibits the longest update time, taking more than 100 seconds to insert a node into a 1-million-node cluster. This high overhead is due to its high complexity. Each node insertion requires $O(m \log(m))$ time, where $m$ is the size of the large lookup table, which is $100\times$ greater than the number of nodes $n$. \par

\ding{173} Ring's update time grows slowly as the cluster expands. When the number of nodes increases from one thousand to one million, the update time changes from 137 us to 455 us. The update overhead of Ring is moderate compared to other schemes, as its update complexity is $O(\log(n))$, which is far less than that of Maglev. \par

\ding{174} AH and DH-IQ exhibit very short update times, each less than 1 us. With a constant update complexity, they can update node states on a nanosecond scale, regardless of the cluster scale. \par

\ding{175} Compared to DH-IQ, DH-B and DH-b have higher insertion overhead. As the active ratio increases from 0.1 to 0.99, the update time of DH-b increases from 2.7 to 245 us, while that of DH-B increases from 1.1 to 80 us. DH-B performs better than DH-b because the byte array has higher efficiency for updates. However, both schemes perform worse than DH-IQ, indicating that the IQueue significantly reduces insertion overhead. \par

Overall, DH-IQ demonstrates the best update performance among the DH variants, while AH also performs impressively. Maglev exhibits the highest update time due to its complex update process, while Ring maintains a moderate update overhead. \par

\begin{figure*}[htb]
  \center{
    \subfloat[Load Balance] {
      \label{fig: balance}
      \includegraphics[width=6.22cm]{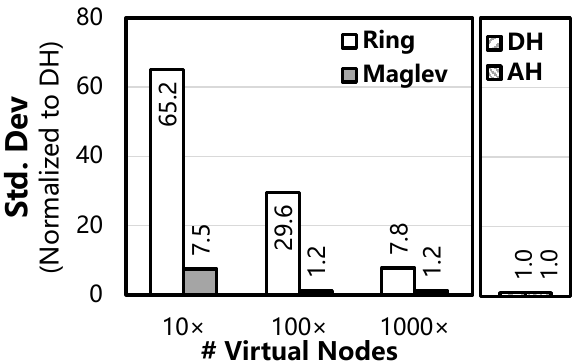}}
    \subfloat[Remapping Ratio] {
      \label{fig: remapping}
      \includegraphics[width=5.82cm]{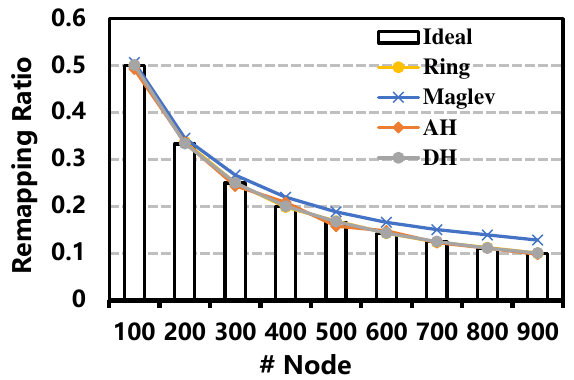}}
  }
\caption{(a) Load balance comparison when the number of virtual nodes per physical node varies from 10 to 1000. The Y-axis is the standard deviation normalized to DH. (b) The remapping ratio after inserting $(9\times 100)$ nodes into a 100-node cluster. Bars represents the ideal remapping ratio, and the dash lines represents real remapping ratio of CH algorithms. }
\label{fig: performance_2}
\end{figure*}

\begin{figure*}[htb]
  \center{
    \subfloat[] {
      \label{fig: ASL}
      \includegraphics[width=4.22cm]{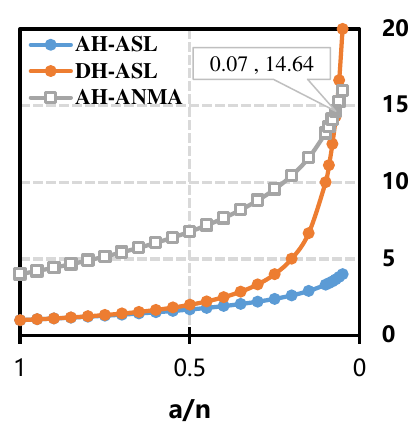}}
    \subfloat[] {
      \label{fig: shrinking}
      \includegraphics[width=4.9cm]{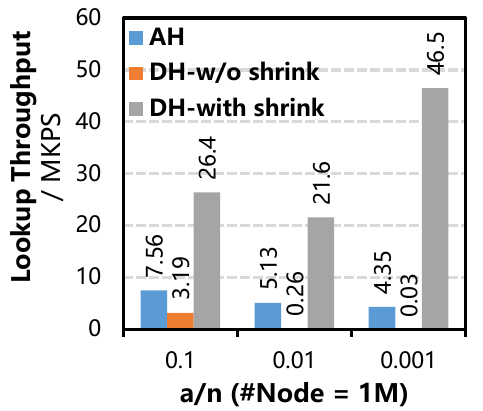}}
    \subfloat[] {
      \label{fig: weighted_test}
      \includegraphics[width=4.67cm]{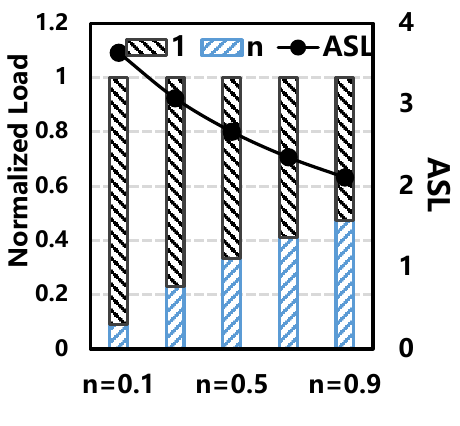}}
  }
\caption{(a) The Average Search Length (ASL) and Average Number of Memory Accesses (ANMA) of querying keys. The X-axis is the active ratio. (b) The lookup performance of AH, DH-B without shrinking, and DH-B with shrinking when the failure ratio is 0.9, 0.99, and 0.999 in 1 million nodes. (c) The ASL (drawn as lines) and load per node (drawn as stacked bars) in weighted DH. The cluster size is 1000, and the load is 100 million lookups. The weights of 512 nodes are 1, and the weights of others are $n$. Here, $n$ is set to 0.1, 0.5, and 0.9 respectively. }
\label{fig: ASL and weighted}
\end{figure*}

\subsection{Balance}
We compare the load balancing performance of Ring, Maglev, AH, and DH. It is worth noting that we do not test DH in different versions because the mentioned optional trade-offs do not affect the load distributions. The cluster consists of 1000 active nodes. Ring's load balancing depends on the design of virtual nodes. Hence, we evaluate Ring with each physical node matching 1, 10, 100, and 1000 virtual nodes, respectively. Maglev's load balancing is influenced by the size of the lookup table, which we set as prime numbers approximately 10 times, 100 times, and 1000 times larger than the node size. Since AH and DH not use over-provisioned memory for balance, they are test once as shown in the right of Figure \ref{fig: balance}. We randomly generate 100 million integers as keys to query the corresponding node IDs. The load on each node is measured by the number of keys assigned to it. In this case, the average load on each node is $100M/1000 = 100K$ queries. We quantify load balance using the standard deviation ($\sigma$). A smaller standard deviation indicates better load balance in a CH algorithm. \par

The experimental results are shown in Figure \ref{fig: balance}. The x-axis represents the number of virtual nodes per physical node, which impacts the load balance of Ring and Maglev. AH and DH are evaluated in a cluster with 1000 active nodes. The y-axis represents the standard deviation of loads in the different CH algorithms. To facilitate comparison, all results are normalized to DH's standard deviation. From the figure, we can observe that all algorithms, except for Ring, achieve good load balance. Ring exhibits the worst load balance, with a standard deviation approximately $7.8\times$ higher than that of DH. These results align with previous research findings \cite{JumpHash}. \par

\subsection{Minimal Disruption}
We evaluate the remapping ratio of Ring, Maglev, AH, and DH after node insertions. We gradually insert 100 nodes into a 100-node cluster until the number of active nodes reaches 1000. The cluster size ranges from 100 to 1000 in increments of 100. As shown in Figure \ref{fig: remapping}, the ideal remapping ratio is calculated by dividing the number of updated nodes by the total number of nodes. The ideal remapping ratios after each insertion are $100/200=0.5$, $100/300=0.33$, $100/400=0.25$, and so on, represented by the bars in the figure. After each insertion, we provide duplicate sets of 100 million keys as input to the CH algorithms and calculate the corresponding node IDs. We compare the current results with the previous insertion's results, count the number of remapped keys, and divide it by 10 million to obtain the real remapping ratio. The real remapping ratios of the different CH algorithms are represented by the dashed lines in Figure \ref{fig: remapping}. Comparing the real remapping ratios to the ideal remapping ratio, we observe that all schemes, except for Maglev, exhibit remapping ratios close to the ideal value. Maglev deviates slightly from the ideal remapping ratio, indicating that it cannot achieve complete minimal disruption, which is consistent with previous research \cite{maglev}. \par

\subsection{Fault Tolerance} \label{sec:EvaFT}

In this section, we test the fault tolerance of AH and DH, as they have demonstrated comparable performance in the previous experiments. DH and AH both require multiple searches to return mapping results, and the search length increases as the node active ratio decreases. We measure their fault tolerance using the average search length (ASL). Initially, the number of nodes is set to 1000, and we gradually remove 100 nodes until only 100 active nodes remain. This procedure reduces the active ratio from 1 to 0.05. Figure \ref{fig: ASL} illustrates the ASL as a function of the active ratio. DH exhibits a larger ASL than AH due to its higher lookup complexity ($O(\frac{n}{a})$) compared to AH's complexity ($O(1+\log(\frac{n}{a})^2)$) \cite{AnchorHash}. However, Section \ref{sec: EV_lookup} demonstrated that the lookup throughput of DH is higher. This is because AH is a stateful algorithm that incurs more memory access overhead to maintain the update order of nodes. AH requires four memory accesses for each search, while DH only requires one. To compare the memory access overhead, we present AH's Average Number of Memory Access (ANMA) for each lookup in Figure \ref{fig: ASL}, where $ANMA_{AH} = 4 \times ASL_{AH}$. The results show that AH consistently has a larger ANMA than DH until the active ratio drops below 7\%. Therefore, when the active ratio is above 7\%, DH consistently outperforms AH in terms of lookup performance, despite its higher lookup complexity. \par

\subsection{Elasticity} \label{sec: EV_scale}

DH supports the operations of scaling and shrinking to adjust the cluster size dynamically. In this section, we specifically evaluate the shrinking operation. We compare three schemes: AH, DH-B without shrink, and DH-B with shrink. The shrinking operation is triggered when the active ratio falls below $0.1$. We measure the lookup throughput of the three schemes in a cluster with 1 million nodes and active ratios of 0.1, 0.01, and 0.001. The experimental results are shown in Figure \ref{fig: shrinking}. From the figure, we make two observations. \par

First, the scheme of DH without shrink performs poorly when the active ratio is low. Compared to AH, which maintains a lookup rate of 4.35 MKPS even at an active ratio of 0.001, DH-B only achieves a lookup rate of 0.03 MKPS. This confirms that DH has a higher lookup complexity than AH. Second, the shrinking operation enhances the elasticity of DH. When shrinking is triggered, DH dynamically adjusts the upper bound of the cluster based on the number of active nodes. As a result, the active ratio increases, significantly improving the lookup throughput. \par

It is worth noting that while scaling and shrinking operations improve lookup performance, they come at the cost of remapping a large number of keys. The remapping ratio depends on the ratio of the cluster size before and after scaling or shrinking. For example, if a cluster shrinks to 1\% of its original size, the remapping ratio is approximately 99\%. If a cluster doubles its size, the remapping ratio is 50\%. Considering the significant remapping overhead, scaling and shrinking operations are best suited for inserting or removing active nodes in batches to amortize the remapping cost. \par

\subsection{Weighted DxHash}

In this section, we evaluate weighted DxHash and verify whether the load distribution aligns with the theoretical derivation. We construct a weighted NSArray consisting of 1024 items, divided into two halves. The weights of one half are uniformly set to 1 (referred to as 1-nodes), while the weights of the other half range from 0.1 to 0.9 in steps of 0.2 (referred to as n-nodes). We generate 10 million random keys as input for weighted DxHash. The loads on each node are measured by the number of keys mapped to that node. Additionally, we record the average search length (ASL) for all keys. Figure \ref{fig: weighted_test} displays the normalized loads on the two parts and the average ASL. The ASL and the load distribution align with Formula \ref{for:weightedASL} and \ref{for:weightedAcc} within an error margin of 0.1\%. This confirms that weighted DxHash effectively adjusts the loads on nodes based on their weights, demonstrating quantitative load distribution control. \par

\section{Conclusions} \label{sec:Con}
This paper introduces DxHash, an efficient, scalable, and adaptable consistent hashing algorithm. We present the algorithm, its implementation, and provide a complexity proof for DxHash. Building on naive DxHash, we propose weighted DxHash. The evaluation of DxHash, compared to other existing CH algorithms, demonstrates its ability to maintain millions of nodes while delivering a high key lookup rate, occupying minimal memory footprint, and requiring minimal time for node additions or removals. Weighted DxHash also achieve its design objectives. Finally, the source code for DxHash and all associated tests are available as open source \footnote{The code is available at https://github.com/ChaosD/DxHash}. \par

\bibliographystyle{plain}
\bibliography{references}

\end{document}